\documentclass[10pt, final, twoside, journal]{IEEEtran}

\usepackage{cite}

\usepackage{amsfonts}
\usepackage{amsmath}
\usepackage{amssymb}
\usepackage{url}

\ifx\pdfoutput\undefined
  \usepackage{graphicx}
\else
  \usepackage[pdftex]{graphicx}
\fi

\newcommand{\est}{{\rm est}}
\newcommand{\sinc}{\ensuremath{{\rm sinc}}}

\newcommand{\hpi}{\ensuremath{\frac{\pi}{2}}}
\newcommand{\kmax}{\ensuremath{k_{\rm max}}}
\newcommand{\kmaxt}{\ensuremath{k_{{\rm max},\theta}}}
\newcommand{\kmaxp}{\ensuremath{k_{{\rm max},\phi}}}
\newcommand{\Kmax}{\ensuremath{K_{\rm max}}}
\newcommand{\hKmax}{\ensuremath{\hat{K}_{\rm max}}}
\newcommand{\hT}{\ensuremath{\hat{\Theta}}}
\newcommand{\Dkp}{\ensuremath{\Delta k_\phi}}
\newcommand{\Dkt}{\ensuremath{\Delta k_\theta}}
\newcommand{\DTest}{\ensuremath{\Delta \Theta_\est}}
\newcommand{\DPest}{\ensuremath{\Delta \Phi_\est}}
\newcommand{\DT}{\ensuremath{\Delta \Theta}}
\newcommand{\Dkr}{\ensuremath{\Delta k_r}}
\newcommand{\FOV}{\ensuremath{FOV}}
\newcommand{\FOVp}{\ensuremath{\FOV_\phi}}
\newcommand{\FOVt}{\ensuremath{\FOV_\theta}}
\newcommand{\res}{\ensuremath{res}}
\newcommand{\rect}{\ensuremath{{\rm rect}}}
\newcommand{\phiwid}{\ensuremath{\phi_{\rm width}}}
\newcommand{\thetawid}{\ensuremath{\theta_{\rm width}}}
\newcommand{\dcf}{\ensuremath{D}}
\newcommand{\dcft}{\ensuremath{\dcf_\theta}}
\newcommand{\dcfp}{\ensuremath{\dcf_\phi}}

\newcommand{\FT}{\ensuremath{\mathcal{FT}}}

\begin{document}

\title{Anisotropic Field-of-Views in Radial Imaging}

\author{Peder~E.~Z.~Larson, Paul~T.~Gurney, and
Dwight~G.~Nishimura,~\IEEEmembership{Member,~IEEE}%
\thanks{This work was supported by NIH grants 2R01-HL39297 and 1R01-EB002524, 
and GE Medical Systems.}%
\thanks{The corresponding author can be reached at peder@mrsrl.stanford.edu.}%
\thanks{The authors are with the Magnetic Resonance Systems Research
  Laboratory, Stanford University, Stanford, CA 94305.}%
\thanks{Copyright (c) 2007 IEEE. Personal use of this material is permitted. However, permission to use this material for any other purposes must be obtained from the IEEE by sending a request to pubs-permissions@ieee.org.}}

\markboth{Transactions on Medical Imaging}{Larson
  \MakeLowercase{\textit{et al.}}: FOVs in Radial Imaging}


\maketitle
\begin{abstract}
Radial imaging techniques, such as projection-reconstruction (PR),
are used in MRI for dynamic imaging, angiography, and
short-$T_2$ imaging.  They are robust to 
flow and motion, have diffuse aliasing patterns, and support short
readouts and echo times.  One drawback is that standard
implementations do not support anisotropic field-of-view shapes, which are
used to match the imaging parameters to the object or region-of-interest.
A set of fast, simple algorithms for
2D and 3D PR, and 3D cones acquisitions are
introduced that match the sampling density in frequency space to the
desired field-of-view shape.
Tailoring the acquisitions allows for reduction of aliasing artifacts in
undersampled applications or scan time reductions without introducing
aliasing in fully-sampled applications.
It also makes possible new radial imaging applications that were
previously unsuitable, such as imaging elongated regions or thin
slabs.
2D PR longitudinal leg images and
thin-slab, single breath-hold 3D PR abdomen images, 
both with isotropic resolution, demonstrate these new possibilities.
No scan volume efficiency is lost by using anisotropic field-of-views.
The acquisition trajectories can be computed on a scan by scan basis.
\end{abstract}




%
%
%
%


\section{Introduction}
\PARstart{R}{adial} medical imaging methods were first used in X-ray
computerized tomography (CT) where data is acquired on radial
projections.  These projections correspond to radial lines in
frequency space, a fact that inspired the first magnetic resonance imaging (MRI)
acquisitions to also occur on radial lines\cite{Lauterbur}.
These 2D imaging methods are also known as projection-reconstruction
(PR) and projection acquisition (PA).
Radial imaging also refers to 3D acquisition techniques such as 3D PR,
as well as 3D cones \cite{IrarrazabalCones, GurneyCones}.

Radial MRI is used for
dynamic imaging applications because it is inherently robust to
motion \cite{GloverPR} and flow \cite{NishimuraFlow}. 
It is used in angiography\cite{PetersPRMRA}, with applications such as contrast-enhanced angiography\cite{BargerVIPR, LuMultiechoPR} 
and time-resolved angiography\cite{VigenPRTRICKS}.
Radial MRI readouts can support very short repetition times (TRs) and
echo times (TEs).  Short TRs are useful for
steady-state free precession (SSFP) imaging, particularly if high resolution
is required or for quadrature fat/water separation.
Both the short TRs and robustness to motion and flow have been taken
advantage of in balanced SSFP coronary artery imaging
\cite{StehningCoronaryFFE, Stehning3DPRNav}.
Short TEs are required for ultra-short echo time (UTE) MRI, which
can image collagen-rich tissues such as tendons, ligaments and menisci, 
as well as calcifications, myelin, periosteum and cortical bone
\cite{GateByd,  RobsonJCAT}.  In addition to 2D acquisitions, 3D
radial imaging techniques, such as 3D PR \cite{Rahmer3DUTE} and 3D
cones \cite{LarsonT2supp}, have been applied to UTE.

3D PR has also been used for the Vastly undersampled Isotropic Projection
Reconstruction (VIPR) imaging method \cite{BargerVIPR} for MR
angiography and efficient phase-contrast flow imaging
\cite{GuPC-VIPR}.  The isotropic resolution of 3D PR is advantageous
for multiplanar and 3D reformats.
VIPR uses undersampled acquisitions because the number of
projections required for fully-sampled 3D PR is prohibitively large.
The undersampling is tolerated
because of the diffuse aliasing properties of radial trajectories \cite{BargerVIPR}.
Another solution to the often prohibitively large number of projections
required by 3D PR is to use conical trajectories \cite{IrarrazabalCones, GurneyCones}.
These sample in a spiral pattern along cones with various projection angles,
allowing for significantly faster volumetric coverage than 3D PR.

In 2D and 3D PR trajectories, as well as 3D cones, a critical number
of projection lines must be
acquired to support a given field-of-view (FOV) -- also referred to as a
region-of-support or region-of-interest.  The lines are normally
acquired with equiangular spacing, resulting in only circular and
spherical FOV shapes.
Many imaging applications, however, have anisotropic dimensions and
would benefit from anisotropic FOVs.

In this paper we introduce a new method to simply and easily design
radial imaging trajectories for anisotropic FOVs.  
One previous method applied several varying angular density functions to
obtain anisotropic 2D FOVs \cite{SchefflerNUPR, SchefflerRFOV}.  
Our method also utilizes non-uniform angular spacing, and is able to exactly
match the sampling to the desired FOV shape.  Design algorithms for 2D and
3D PR, as well as 3D cones are presented.
Tailoring the FOV for non-circular objects or regions-of-interest
allows for scan time reductions without introducing aliasing
artifacts.  In undersampled applications, this tailoring will reduce
the occurrence of aliasing artifacts.
Additionally, our algorithms support anisotropic
projection lengths, and, equivalently, resolution.
This is useful for reducing gradient slew rate
demands in 3D cones, and could also be used for reducing
gradient distortion artifacts by favoring the more homogeneous or
powerful axes.
It can also be used to trade image resolution in certain dimensions
for acquisition time.


\section{Theory}
In radial imaging, the raw data is acquired on radial lines in
frequency space, referred to as ``k-space'' in MRI.  
A discrete number of these lines, also known as projections or spokes,
are sampled during data acquisition.
Sampling theory tells us that the sample separation determines the FOV. 
Cartesian samples are generally equally spaced, and the
FOV is equal to the inverse of the sample spacing, $\frac{1}{\Delta k}$.
In radial imaging, the spacing between adjacent projections in k-space varies
radially, and is most sparse at the end of the
projections.  The sample spacing along the projections must also be
considered, although in MRI this usually does not limit the FOV as much as the
distance between projections.
We will investigate how this projection separation determines the FOV
and resolution.  Further discussion
of radial sampling theory can be found in~\cite{LauzonPolarSampling}.

\subsection{FOV vs. PSF}
\label{sec:fov_vs_psf}
The FOV of a given sampling pattern is characterized by its Fourier
transform (\FT), known as the point spread function (PSF).
Since the sampling pattern is multiplied by the frequency data,
the PSF is convolved with the image data.
The central lobe of the PSF introduces a blurring, and thus determines
the resolution of the resulting image.
Aliasing lobes in the PSF are outside this central lobe and determine the
FOV of the resulting image.  In radial imaging, the distinct shape of
the aliasing lobe determines the FOV, but they are not always the same
shape.
Convexity of the aliasing lobe shape in the PSF will ensure that the
FOV shape is exactly the same.
Polygons, rectangles, cuboids, ellipsoids, and ovals are some examples
of convex shapes useful for medical imaging.
While there are concave PSF shapes that will support concave FOVs,
such as a ``plus''-sign shape,
we will assume convexity since this includes most objects and
regions-of-interest in medical imaging.

\begin{figure}
  \begin{center}
    \includegraphics[width=\columnwidth]{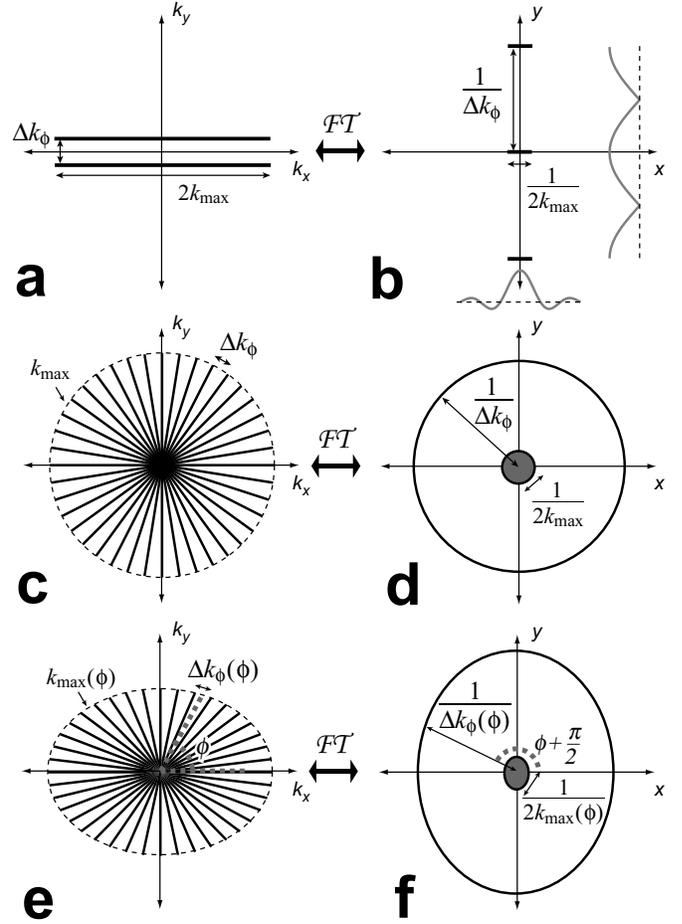}
    \caption{
      Sampling in 2D radial imaging.
      (a) Approximation of adjacent projections by parallel sampled lines.
      (b) Parallel lines PSF illustration.  The gray lines show the
      spatial variation derived in Eq.~\ref{eq:psflines}.  Aliasing
      lobes appear at $\frac{1}{\Dkp}$.  With additional projections,
      the aliasing lobes become much sharper that the cosine shown.
      (c) Isotropic PR trajectory with constant angular spacing and extent.
      (d) Isotropic PR PSF illustration.
      (e) PR trajectory with variable angular spacing and extent.
      (f) Angularly varying PR PSF illustration.
    }
    \label{fig:sampling_approx}
  \end{center}
\end{figure}

\subsection{2D Sampling}
The FOV of a 2D radial imaging trajectory can be analyzed by
decomposition into adjacent spokes.  A pair of adjacent spokes can be
approximated as two parallel lines, shown in
Fig.~\ref{fig:sampling_approx}a and described mathematically as
\begin{equation}
\label{eq:paralines}P(k_x,k_y) = [ \delta(k_y - \frac{\Dkp}{2}) + \delta(k_y + \frac{\Dkp}{2}) ] \cdot
\rect(\frac{k_x}{2\kmax}),
\end{equation}
where $\Dkp$ is the separation at the end of the projections and $\kmax$
is the length of the projections.
The PSF of Eq.~\ref{eq:paralines}, illustrated in
Fig.~\ref{fig:sampling_approx}b, is: 
\begin{equation}
\label{eq:psflines}\FT\{P(k_x,k_y)\} = C \cos(\pi \Dkp y) \cdot \sinc(2\kmax x),
\end{equation}
where $C$ includes all scaling factors.  The peaks of 
$| \cos(\pi \Dkp y) |$ will introduce aliasing when convolved with the
object, thus limiting the FOV perpendicularly to the
direction of the lines to $\frac{1}{\Dkp}$.  
The actual PSF is much more peaked than a cosine because
the sample spacings in $k_y$ from other
projections introduce phase variations that cause the aliasing to cancel
out along $y$ between $0$ and $\pm\frac{1}{\Dkp}$, as is shown in the
``Results'' section.  This is analogous
to Cartesian sampling where the PSF of two parallel lines is also a cosine
but phase variations from the other sampling lines
cause sharp aliasing peaks.
The resolution is limited to a minimum of $\frac{1}{2\kmax}$ due the
$\sinc(2\kmax x)$ term.  

If all projections are equally spaced, as shown in
Fig.~\ref{fig:sampling_approx}c, the
alias-free FOV and resolution, \res, are:
\begin{eqnarray}
\FOV &=& \frac{1}{\Dkp} \approx \frac{1}{\kmax\DT} =  \frac{N}{\pi\kmax}, \\
\res &=& \frac{1}{2\kmax},
\end{eqnarray}
where $\DT$ is the angular spacing between spokes in radians, $N$
is the number of full projections acquired, and $\kmax$
is the extent of the spokes in k-space (Fig.~\ref{fig:sampling_approx}d).
Here we have used the approximation that $\sin(\DT) \approx \DT$
for small \DT.

From Fourier theory, rotation of the parallel lines rotates their \FT, thus
the angular spacing between spokes can be varied to produce
anisotropic FOV shapes.
Similarly, if $\kmax$ is varied, the resolution size changes angularly.
Equation~\ref{eq:psflines} tells us that the sample spacing at angle $\phi$
determines the perpendicular FOV, while $\kmax$
at $\phi$ determines the resolution at $\phi$. 
This is shown in Fig.~\ref{fig:sampling_approx}e and f, and is formally expressed as,
\begin{eqnarray}
\label{eq:vFOV} \FOV(\phi+\hpi)& =& \frac{1}{\Dkp(\phi)} \approx
\frac{1}{\kmax(\phi)\DT(\phi)}, \\
\label{eq:vres} \res(\phi)& =& \frac{1}{2\kmax(\phi)}.
\end{eqnarray}

Discrete sampling along the parallel lines results in repetition of
the \FT~pattern (Eq.~\ref{eq:psflines}) in this sampling direction ($x$
in Eq.~\ref{eq:psflines} and Fig.~\ref{fig:sampling_approx}a).  Radial
sample spacing of \Dkr~results in repetitions at $\frac{1}{\Dkr}$.  
These aliasing lobes are often eliminated in MRI because of low-pass filters
applied during the readout.
These filters will, however, limit the FOV in the projection
direction to $\frac{1}{\Dkr}$ because they are matched to the sample
spacing.  
Therefore, the maximum radial sampling spacing must be
\begin{equation}
\label{eq:maxspace} \Dkr \le \frac{1}{\max(\FOV(\phi))}
\end{equation}
to ensure there is no aliasing or FOV restrictions due to the
sampling along the projections.

\subsection{3D Sampling}
Radial sampling in 3D can be analyzed with the same principles as in 2D.
Again, approximating
adjacent spokes as a set of parallel lines, we now have:
\begin{eqnarray}
P(k_x,k_y,k_z) & = &  [ \delta(k_y - \frac{\Dkp}{2}) + \delta(k_y +
  \frac{\Dkp}{2}) ] \cdot \nonumber \\
\label{eq:paralines3} & & \rect(\frac{k_x}{2\kmax}) \cdot \delta(k_z).
\end{eqnarray}
The PSF is then the same as Eq.~\ref{eq:psflines}:
\begin{equation}
\label{eq:psflines3}\FT\{P(k_x,k_y,k_z)\} = K \cos(\pi \Dkp y) \cdot \sinc(2\kmax x).
\end{equation}
This aliasing pattern is the same as illustrated in
Fig.~\ref{fig:sampling_approx}b, but with an infinite extent in $z$. 

\begin{figure}
  \begin{center}
    \includegraphics[width=\columnwidth]{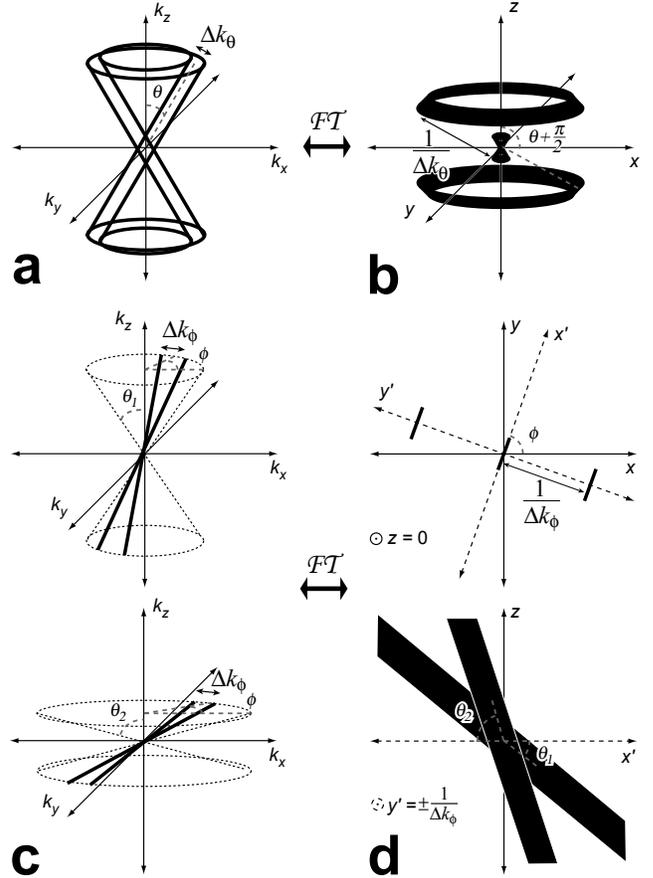}
    \caption{
      Sampling in 3D radial imaging.
      (a) Approximation of adjacent cones by rotated parallel lines.
      (b) PSF illustration of adjacent cones model, which is the
      parallel lines PSF (Fig.~\ref{fig:sampling_approx}b) rotated
      around the $z$-axis.
      (c) Adjacent projections on two different cones.
      (d) PSF illustrations for the adjacent projections.
      The top illustration shows the intersection of the PSF with the
      $x$-$y$ plane, which is approximately identical for the
      projections on both cones. 
      The bottom illustration shows the planes which contain the
      aliasing lobes.  These planes are perpendicular to
      the $y'$ axis at $y' = \pm 1/\Dkp$, and the lobes extend and
      different angles depending on the cone angle.
    }
    \label{fig:3Dsampling_approx}
  \end{center}
\end{figure}

A set of adjacent cones can be modeled by rotating a pair of
parallel lines about the $k_z$-axis, as shown in
Fig.~\ref{fig:3Dsampling_approx}a, which rotates the PSF of these parallel
lines about the $z$-axis.  The resulting PSF pattern
is circularly symmetric about the $z$-axis and with polar
deflection of the aliasing lobe of $\theta + \hpi$ for a cone
deflection of $\theta$, shown in Fig.~\ref{fig:3Dsampling_approx}b.
We are using the spherical coordinate notation where $\theta$, the polar angle,
is the deflection from the positive $z$-axis and $\phi$, the azimuthal angle,
is the rotation from the positive $x$-axis.

Adjacent projections on a cone produce aliasing patterns
that are rotations of the PSF in Eq.~\ref{eq:psflines3}, as 
illustrated in Fig.~\ref{fig:3Dsampling_approx}c and d.
The aliasing lobes are the nearest to the origin in the $x$-$y$ plane
at an azimuthal angle $\phi+\hpi$
and at a distance of $1/\Dkp$.
They extend in a direction that is determined by the cone deflection 
angle, as shown in Fig.~\ref{fig:3Dsampling_approx}d.

Combining these results, we find that the sampling along a cone limits
the FOV azimuthally in the $x$-$y$ plane while the sampling of different
cones limits the FOV in the polar direction.  Mathematically, 
\begin{eqnarray}
\label{eq:FOVtheta} \FOV_\theta(\theta + \hpi,\phi)& =& \frac{1}{\Dkt(\theta)} \\
\label{eq:FOVphi} \FOV_\phi(\theta = \hpi, \phi + \hpi)& =& \frac{1}{\Dkp(\phi)} \\
\label{eq:res3D} \res(\theta, \phi)& =& \frac{1}{2\kmax(\theta,\phi)},
\end{eqnarray}
where the variables are all illustrated in Fig.~\ref{fig:3Dsampling_approx}.
The resolution is determined by the projection extents.
The samples along the projection must again satisfy Eq.~\ref{eq:maxspace}.


\section{Methods}

\subsection{2D Projection-Reconstruction}
Two-dimensional radial imaging trajectories can be defined by the
projection angles, projection lengths, and the sampling pattern along
the projections.  Our algorithm designs a set of projection angles, $\Theta[n]$,
and projection lengths, $\Kmax[n]$, for a desired FOV
pattern.  The sampling pattern along the projections is not designed,
and should be chosen to meet the condition in Eq.~\ref{eq:maxspace}.

\begin{figure}
\begin{center}
\framebox{
\begin{minipage}[t]{.95\columnwidth}
  \newcounter{Lcount}
  \begin{list}{\arabic{Lcount}.}{\usecounter{Lcount}
      \setlength{\leftmargin}{5pt} }
  \item  Initialize $\Theta[1] = \phi_0$ and $n = 1$

  \item \label{2Dalg:calc} Calculate:
    \begin{gather*}
      \DTest = \frac{\textstyle 1}{\textstyle \kmax(\Theta[n]) \FOV(\Theta[n] + \hpi)} \\
      \DT  =  \frac{\textstyle 1}{\textstyle \kmax(\Theta[n] + \frac{\DTest}{2})
                  \FOV(\Theta[n] +\frac{\DTest}{2} + \hpi)} \\
      \Theta[n+1]  =  \Theta[n] +  \DT 
    \end{gather*}
    and increment $n$

  \item Repeat previous step until $\Theta[n] > \phi_0 + \phiwid$
    
  \item \label{2Dalg:choose} Choose a scaling factor, $S$, and the
    number of projections returned, $N$, as follows: \\
    If $\Theta[n] - (\phi_0 + \phiwid)  < (\phi_0 + \phiwid) - \Theta[n-1] $ \\
    \hspace*{15pt}choose $S = \frac{\textstyle \phiwid}{\textstyle \Theta[n]-\phi_0}$ and $N = n-1 $ \\
    Else \\
    \hspace*{15pt}$S = \frac{\textstyle \phiwid}{\textstyle \Theta[n-1]-\phi_0}$ and $N = n-2 $ \\

 \item \label{2Dalg:scale} Scale the set of angles \\
   $\Theta[1, \ldots, N] = S \cdot (\Theta[1, \ldots, N] - \phi_0) + \phi_0$

 \item \label{2Dalg:kmax} Calculate $\Kmax[n] = \kmax(\Theta[n])$

  \end{list}

\end{minipage}
}
\end{center}
\caption{Generalized 2D anisotropic FOV radial imaging algorithm}
\label{fig:2Dalg}
\end{figure}

The algorithm is shown in Fig.~\ref{fig:2Dalg}.  
The desired FOV must be specified as a function of angle, $\FOV(\phi)$,
and must be $\pi$-periodic ($\FOV(\phi) = \FOV(\phi + \pi)$).  The
desired projection lengths can be specified as $\kmax(\phi)$, and also
must be $\pi$-periodic.
An initial angle of $\phi_0$ may be specified, although the default of
$\phi_0 = 0$ is appropriate for most 
applications.  The angular width of the resulting angles, \phiwid, may
also be specified, and will be $\pi$ for most applications,
including full-projection PR.  One application of varying this width
is half-projection PR, where $\phiwid = 2\pi$ could be used.  These parameters are
also useful for designing 3D PR trajectories - see Section~\ref{sec:3DPR}.

After initialization, the relationship derived in
Eq.~\ref{eq:vFOV} is used to sequentially calculate a set of projection
angles in step~\ref{2Dalg:calc} of the algorithm.  The projection
separation is first estimated, \DTest, and then the actual projection
separation is then calculated using the angle $\Theta[n] + \frac{\DTest}{2}$.
Without this estimation, the resulting FOV will be very slightly rotated and
distorted because the projection separation is centered between
$\Theta[n]$ and $\Theta[n+1]$.  Since $\Theta[n+1]$ is unknown,
$\Theta[n] + \frac{\DTest}{2}$ provides a good estimate of the middle angle,
especially compared to using $\Theta[n]$.
Additional iterations for increased accuracy are possible
but provide little additional benefit.

The sequential nature of the algorithm and the required periodicity of
the projection angles results in 
an undesired angle spacing at the end of the design.
Steps~\ref{2Dalg:choose} and~\ref{2Dalg:scale} correct for this 
by scaling the set of projection angles to the chosen \phiwid.  
The scaling slightly distorts the resulting FOV shape so  
the scaling factor, $S$, is chosen by step~\ref{2Dalg:choose} to be as
close to 1 as possible.
This step could also be modified to require that
the number of projections be even, odd, or a scalar multiple 
for acquisition strategies such as multi-echo
PR~\cite{LuMultiechoPR}.

The computation cost of this algorithm is small since it
involves simple calculations and requires no iterations or large matrix operations.

\subsection{3D Cones}
\begin{figure}[b!]
  \begin{center}
    \includegraphics[width=\columnwidth]{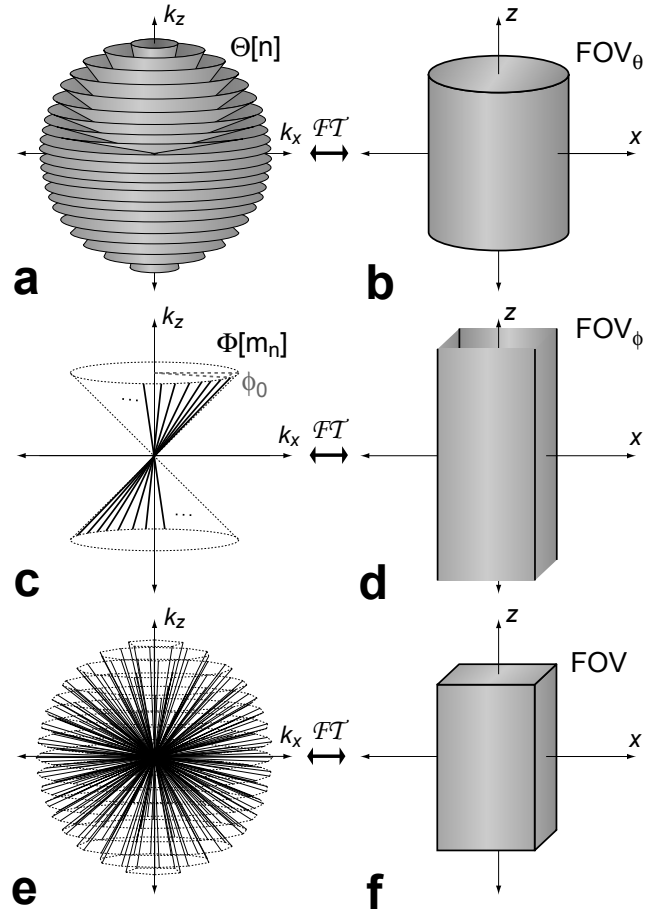}
    \caption{
      3D radial trajectory design (cones-based).
      (a) Cones designed with 2D algorithm.
      (b) Sampling of cones introduces $\FOVt$, which is circularly
      symmetric about the $z$-axis.
      (c) Projections on each cone designed with 2D algorithm using a
      random starting angle $\phi_0$.
      (d) Sampling along cones introduces $\FOVp$, which is
      approximately independent of $z$. 
      (e) 3D PR trajectory resulting from sampling in (a) and (c). 
      (f) The 3D PR FOV is the minimum of $\FOVt$ (b) and
      $\FOVp$ (d).
    }
    \label{fig:3DPR_design}
  \end{center}
\end{figure}

The 2D design algorithm can directly be applied to designing 3D cones
imaging trajectories.  The 3D FOV shapes that are achievable
are circularly symmetric about one axis, which we define to be
the $z$-axis, corresponding to cones wrapping around the $k_z$-axis as 
shown in Fig.~\ref{fig:3Dsampling_approx}a and b.
These shapes can be described by $\FOV(\theta, \phi) = \FOVt(\theta)$,
as is illustrated in Fig.~\ref{fig:3DPR_design}a and b. 
For example, a rectangular-shaped $\FOVt(\theta)$ will yield a cylindrical FOV,
while an elliptical $\FOV(\theta)$ yields an ellipsoid.

For anisotropic 3D cones design, the 2D PR algorithm is used with
inputs of $\FOV(\theta)$ and $\kmax(\theta)$ (optional), $\phiwid =
\pi$, and $\phi_0 = \frac{1}{2\kmax(0)\FOV(\hpi)}$.  
This $\phi_0$ ensures that no cones are just a single projection.
The resulting set of projection angles and extents describe the set of
cones to be used, where the angles, $\Theta[n]$, represent the deflection from the
$k_z$-axis.  Sampling within each cone and generation of appropriate
gradient waveforms are described in~\cite{GurneyCones}.

\subsection{3D Projection-Reconstruction}
\label{sec:3DPR}
\begin{figure}[b]
  \begin{center}
    \includegraphics[width=.9\columnwidth]{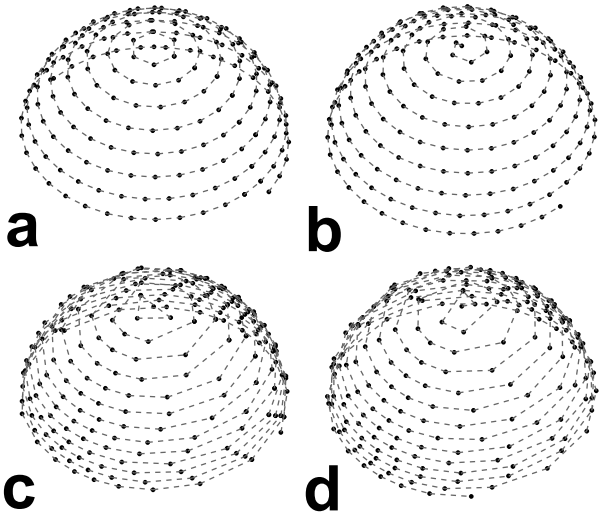}
    \caption{
      3D PR trajectories.
      (a) Cones-based design, and
      (b) Spiral-based design
      for uniform sampling, resulting in a spherical FOV.
      (c) Cones-based design, and
      (d) Spiral-based design
      of an ellipsoid FOV with, non-uniform sampling
      in both the azimuthal and polar directions.
    }
    \label{fig:cones_vs_spiral}
  \end{center}
\end{figure}

We have created two methods for designing 3D PR trajectories with
anisotropic FOVs.  One samples a set of cones, the ``cones-based''
method, while the other designs and samples a spiraling path, the
``spiral-based'' method. 
Figure~\ref{fig:cones_vs_spiral} shows two resulting sampling patterns
for both the cones-based and spiral-based design methods. 

\subsubsection{Cones-based Design}
This method designs 3D projections for an anisotropic FOV by first
designing a set of cones and then appropriately sampling on each cone.
Isotropic sampling in these two dimensions will result in a
spherical FOV, with the number of projections per cone proportional to 
$\sin(\theta)$~\cite{GloverBoron}, where $\theta$ is the polar angle
of the cone.

Equations~\ref{eq:FOVtheta} and~\ref{eq:FOVphi}, illustrated in
Fig.~\ref{fig:3Dsampling_approx}, tell us that the cones sampled
will define a limiting FOV that is circularly
symmetric about the $z$-axis, $\FOVt(\theta,\phi) = \FOVt(\theta)$,
while the samples on each cone define a
maximum FOV that is approximately invariant in $z$, 
$\FOVp(\theta, \phi) \approx \FOVp(x,y)$ (Fig.~\ref{fig:3DPR_design}a-d).
We will describe $\FOVp$  using the deflection angle in cylindrical
coordinates, $\phi_c$, resulting in the simple notation $\FOVp(x,y) = \FOVp(\phi_c)$.
Both sampling patterns limit the FOV, resulting in
\begin{equation}
\label{eq:3DFOV} \FOV =
\min ( \FOVt(\theta), \FOVp(\phi_c) ),
\end{equation}
as illustrated in Fig.~\ref{fig:3DPR_design}e and f.

This method takes inputs of $\FOVt$ and $\FOVp$, with 
\begin{equation}
\label{eq:FOVp}\FOVp(\phi_c) \le \FOVt(\hpi)
\end{equation}
to ensure that the cone spacing does not introduce
aliasing inside $\FOVp$ in the $x$-$y$ plane. 
$\kmaxt(\theta)$ can also be used.  
Varying $\kmaxp$
has not been incorporated into our algorithm
because it causes the polar angle spacing to vary azimuthally,
resulting in distortion of the supported FOV.
The cones are designed using the 2D PR algorithm
with $\FOVt(\theta)$, $\kmaxt(\theta)$, $\theta_0 = 0$, and $\thetawid = \hpi$.
The resulting $\Theta[n]$ and $\Kmax[n]$, where $n = 1, \ldots,
N_{cones}$, are then sampled, also using the 2D PR design algorithm.  

The spokes covering each cone are designed with $\FOVp(\phi_c)$ and $\phiwid = 2\pi$.  
For cone $n$, $\kmaxp = \Kmax[n] \cdot \sin(\Theta[n])$ to adjust for
the circumference of the cone.
The initial angle $\phi_0$ on each cone is chosen at random
uniformly in the interval $[0, \frac{1}{\Kmax[n] \cdot \sin(\Theta[n]) \cdot 
\FOV_\phi(\hpi)}]$.  This randomization reduces coherent aliasing
artifacts that are introduced when $\phi_0=0$ for each cone (see Fig.~\ref{fig:coherent_aliasing}).
Note that the first cone ($n=1$) is just a single projection along the $k_z$-axis.
The result is a set of $\Phi[m_n]$ for $m_n = 1, \ldots, M_n$ 
and $n = 1, \ldots, N_{cones}$, where
$M_n$ is the number of projections within cone $n$.

To complete the full-projection design, an additional cone is added in
the $k_x$-$k_y$ plane ($\theta = \hpi$) that is
only azimuthally sampled along half of the circumference.   
The 2D PR design algorithm is
used, with $\FOVp(\phi_c)$, $\kmaxp = \kmaxt(\hpi)$, a randomized
$\phi_0$, and $\phiwid = \pi$. 
For half-projection designs, the additional half-sampled cone is not needed and
$\thetawid = \pi$ is used for the cone design.


\subsubsection{Spiral-based Design}
This method of 3D PR design is inspired by a method that isotropically samples  
the unit sphere on a spiral path \cite{WongRoos}.  
The design is similar to the cones-based method, taking the same
inputs of $\FOVt$ and $\FOVp$, with $\FOVt(\hpi) =
{\rm max}(\FOVp(\phi_c))$, and, optionally, $\kmaxt$.
This method is best suited for half-projections, and an extension to 
full-projection design is described in Appendix~\ref{app:spiral_full}.
It also results in a more diffuse aliasing pattern 
(see Fig.~\ref{fig:coherent_aliasing}).

First, a set of polar sampling angles is designed with the 2D PR
algorithm exactly as in the cones-based method for half-projections, with $\FOVt$,
$\kmaxt$, $\theta_0 = 0$, and $\thetawid = \pi$, yielding $\hT[n]$ and
$\hKmax[n]$ for $n = 1, \ldots, N_{polar}$.  To create a continuous
sample sample path in the longitudinal direction, these sets are
linearly interpolated to final samples of $\hT[N_{polar}+1]
= \pi$ and $\hKmax[N_{polar}+1] = \kmaxt(\pi)$. 

The interpolation uses the number of projections required, which is
found by estimation using the required number of azimuthal samples on a cone
at $\theta=\hpi$, $N_{\phi,\est}$, as well as $\hT[n]$ and $\hKmax[n]$.
The 2D PR algorithm with $\FOVp$, $\kmaxp = \kmaxt(\hpi)$, $\phi_0 =
0$, and $\phiwid = 2\pi$ is used to compute $N_{\phi,\est}$.
The number of projections between $\hT[n]$ and $\hT[n+1]$ will be
approximately
\begin{eqnarray}
N_{n,\est} &=& N_{\phi,\est} \sin(\frac{\hT[n] +\hT[n+1]}{2}) \nonumber \\
& & \times \frac{\hKmax[n] +\hKmax[n+1]}{2\kmaxp}.
\end{eqnarray}
The total number of projections is chosen to be
\begin{equation}
N = \sum_{n=1}^{N_{polar}} N_{n,\est}.
\end{equation}

The linear interpolation is done such that
there are $N_{n,\est}$ projections between $\hT[n]$ and $\hT[n+1]$.
This is done by using a parametrization, $\theta(t)$, where
$\theta(t_n) = \hT[n]$, $t_{n+1} - t_n = N_{n,\est}$, and $t_1 = 1$.
The polar projection angles are computed by sampling
the linear interpolation of this parametrization as $\Theta[m] =
\theta(m)$ for $m = 1, \ldots, N$.  If variable extents are used, the interpolation of
$k(t)$, where $k(t_n) = \hKmax[n]$, is sampled as $\Kmax[m] = k(m)$.

The azimuthal sampling, $\Phi[m]$, is then computed similarly to
step~\ref{2Dalg:calc} in Fig.~\ref{fig:2Dalg}, with an additional scaling
factor also used in the cones-based design to compensate the cone circumference:
\begin{gather}
  \DPest = \frac{1}{\Kmax[m]\sin(\Theta[m]) \FOVp(\Phi[m] + \hpi)} \\
  \Delta \Phi  =   \frac{1}{\Kmax[m]\sin(\Theta[m])
  \FOVp(\Phi[m] + \frac{\DPest}{2} + \hpi) } \\
  \Phi[m+1]  =  \Phi[m] + \Delta \Phi,
\end{gather}
with the initial angle of $\Phi[1] = 0$.
This results in a set of projection angles, $\Phi[m]$, $\Theta[m]$,
and extents, $\Kmax[m]$ yielding the anisotropic FOV described in Eq.~\ref{eq:3DFOV}.

\subsection{Reconstruction} 
All images were reconstructed with the gridding algorithm
\cite{BeattyGridding}, and PSFs were calculated by gridding data of all ones.
Each data point is multiplied by a density compensation factor (dcf)
to correct for the unequal sample spacing, which results in coloring of the noise
and an intrinsic loss in the signal-to-noise ratio (SNR) efficiency~\cite{TsaiVdens}.
The dcf can be separated into a radial and angular components, where
the angular component accounts for the anisotropic projection spacing.
The angular dcf is chosen so that the same radial dcf can be used
on each projection, provided they all have an equal number of identically
distributed radial samples.
When the samples are equally spaced
this radial component is linear for 2D PR and quadratic for 3D PR.

The 2D angular dcf (\dcft) is proportional to the
projection separation, \Dkp, given from Eq.~\ref{eq:vFOV}.
For a separable dcf,
$\dcft$ must also be proportional to the projection length
because the scaling of the radial sample spacing by $\kmax$ cannot be
described in a single radial dcf.
The resulting 2D PR angular dcf is
\begin{equation}
\label{eq:2d_dcf} \dcft(\Theta[n]) = \Kmax[n] \cdot \Dkp = \frac{\Kmax[n]}{\FOV(\Theta[n] + \hpi)}.
\end{equation}

For anisotropic 3D cones, this compensation factor is applied to each
cone by dividing Eq. 10 in reference~\cite{GurneyCones} by \dcft.
In 3D PR, the density compensation for a given projection is found
by multiplying the compensation factors for the polar and
azimuthal sampling:  
\begin{multline}
\label{eq:3d_dcf}
\dcf_{3D}(\Theta[n],\Phi[m_n]) = \dcft(\Theta[n])\times
\dcfp(\Phi[m_n]) \\
 = \frac{\Kmax[n]}{\FOVt(\Theta[n] + \hpi) \times \FOVp(\Phi[m_n] + \hpi)}.
\end{multline}
For the full-projection spiral-based design, this 
dcf is slightly modified as described in Appendix~\ref{app:spiral_full}.

\subsection{Design Functions}
The anisotropic FOVs were designed in Matlab 7.0 
(The Mathworks, Natick, MA, USA).  The design functions for 2D PR, 3D
cones, and both 3D PR methods, along with
accompanying documentation are available for general use at
\url{http://www-mrsrl.stanford.edu/~peder/radial_fovs}.

\begin{figure}[b!]
  \begin{center}
    \includegraphics[width=\columnwidth]{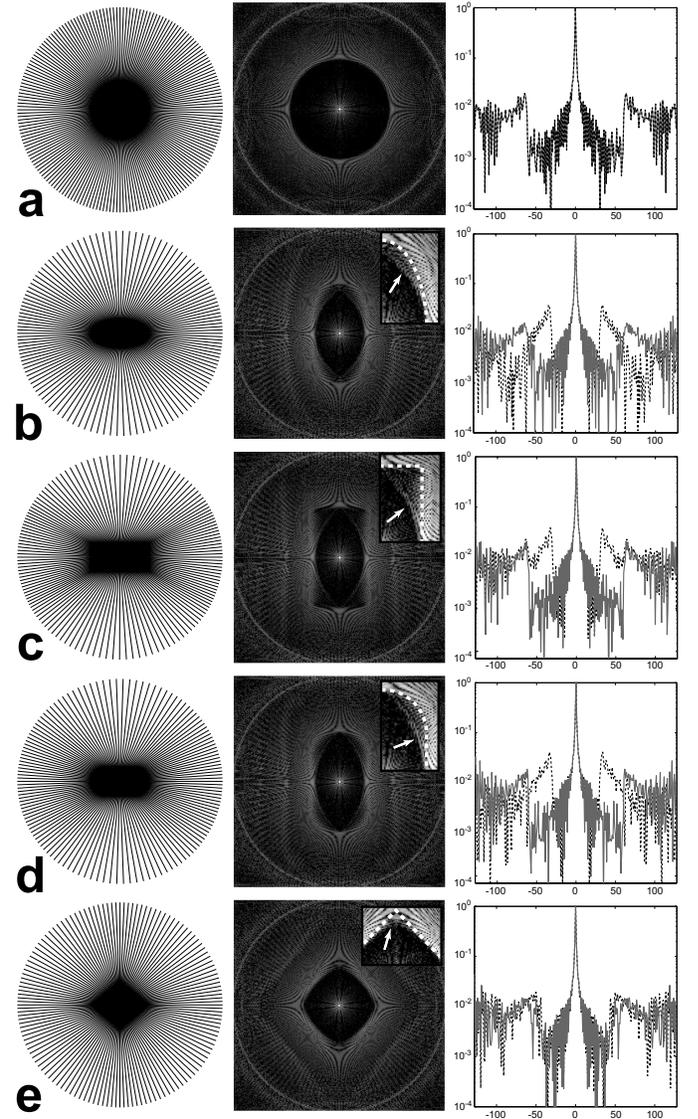}
    \caption{
      Projection sampling patterns (left column) and PSFs (middle
      column), with plots along $x$ (black, dashed line) and $y$ (gray
      line) axes in the right column.
      FOV shapes: (a) Circle, (b) Ellipse, (c) Rectangle, (d) Oval,
      and (e) Diamond.
      The inset PSF images are windowed narrowly to show the low-level
      aliasing (arrows) within the desired FOV (dashed lines).
      The plots show the aliasing peaks and the isotropic resolution
      in the central lobes.
      Small FOVs are used for visualization of the variable angular density.
    }
    \label{fig:2Dpsfs}
  \end{center}
\end{figure}

\subsection{MRI Experiments}
A GE Excite 1.5T scanner with gradients capable of 40 mT/m amplitude and
150 T/m/s slew rate (GE Healthcare, Milwaukee, WI) was used for all
experiments.
The 2D PR images were acquired with a UTE sequence using half-projection acquisitions, 5 mm slice thickness, TE = 500 $\mu$s, 
TR = 100 ms, 30$^\circ$ flip angle, 512 samples per projection, and 
1 mm resolution with a transmit/receive extremity coil.  
The 3D PR images were acquired with a spoiled gradient-recalled echo (SPGR)
sequence using full-projection acquisitions with TE = 3 ms and TR = 10 ms.
The bottle phantom images used a 15$^\circ$ flip angle, 256 samples per projection, and 1 mm isotropic resolution with a transmit/receive head coil.
The in vivo images used a
30$^\circ$ flip angle, 3 cm slab-selective RF excitation, 
192 samples per projection, a body coil, 
and were acquired in a single 25 second breath-hold.
Both fully-sampled and undersampled cylindrical FOVs with 3 and 2 mm
isotropic resolution, respectively, are compared to
isotropic FOVs requiring the same number of projections. 



\section{Results}

\subsection{2D PSFs}
Figure~\ref{fig:2Dpsfs} shows some sampling patterns designed with the
2D anisotropic FOV algorithm and their corresponding PSFs, showing
that the desired FOV shapes are achieved.  They have isotropic resolution,
as shown in the PSF central lobes.
There is some low-level aliasing introduced inside the desired FOV for
the anisotropic shapes, shown in the inset PSF images.
See Section~\ref{sec:discussion} for a full discussion.

\begin{figure}[b!]
  \begin{center}
    \includegraphics[width=\columnwidth]{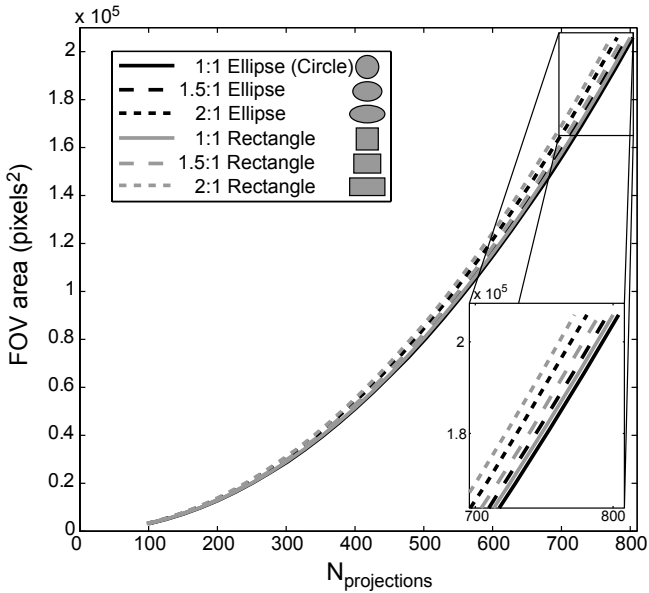}
    \caption{
      Number of projections for elliptical and rectangular FOV shapes versus the resulting
      FOV (shape) area.  All the FOVs have 1 pixel isotropic resolution.
      There is no loss in efficiency by using anisotropic shapes, and
      in fact they are slightly more efficient than the isotropic
      case.  This increase in efficiency comes at the cost of
      low-level aliasing within the FOV, seen in Fig.~\ref{fig:2Dpsfs}.
    }
    \label{fig:area}
  \end{center}
\end{figure}

Figure~\ref{fig:area} shows the relationship between the number of
projections and the FOV area for elliptical and rectangular FOV
shapes.  The relationship is quadratic for a circular FOV,
and the other shapes are also approximately quadratic.  The
anisotropic shapes are slightly more efficient, with efficiency
increasing as the shapes become narrower.
This is due to the low-level aliasing seen in
Fig.~\ref{fig:2Dpsfs}, which is larger for narrower shapes (see Section~\ref{sec:discussion}).

\begin{figure}[h!]
  \begin{center}
    \includegraphics[width=\columnwidth]{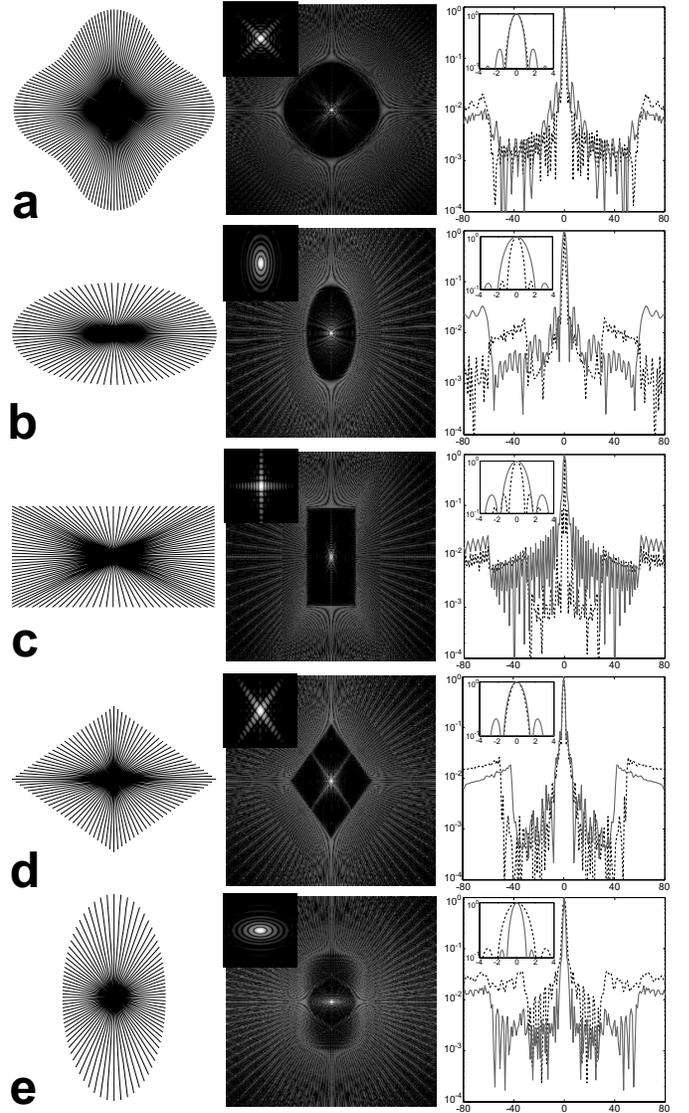}
    \caption{
      Variable $\kmax$ sampling patterns (left column) and
      PSFs (middle column, plots in right column), with enlarged views
      of the central lobe inset.  The black, dashed line plot is along the $x$-axis,
      and the gray line is along the $y$-axis in (b), (c), and (e), and
      along $x=y$ in (a) and (d).
      FOV/\kmax { }shapes: (a) Circle/Star, (b) Ellipse/Ellipse, (c)
      Rectangle/Rectangle, (d) Diamond/Diamond, and (e) Oval/Ellipse.
      All the combinations have a minimum resolution size
      of 1 pixel, with maximum resolutions 2 pixels in (a), (b), and
      (e), 1.72 pixels in (c), and 1.33 pixels in (d).
      The shape combinations in (b), (c), and (d) have no
      low-level aliasing inside the FOV shape, unlike (a) and (e), and
      the isotropic resolution projections in Fig.~\ref{fig:2Dpsfs}.
      Small FOVs are used for visibility of the variable angular density.
    }
    \label{fig:2Dpsfs_kmax}
  \end{center}
\end{figure}

Some sampling patterns with variable $\kmax$ patterns and their
corresponding PSFs are shown in Figure~\ref{fig:2Dpsfs_kmax}.
The angularly varying projection lengths results in anisotropic resolution, seen
in the main lobe of the PSFs.
There is again some low-level aliasing introduced inside the desired
FOV in Fig.~\ref{fig:2Dpsfs_kmax}a and e, but there is none of this
aliasing in b, c, and d.  For these three sampling patterns, $\kmax$ is
the dual of the FOV shape, or $\kmax(\phi) \propto \FOV(\phi+\hpi)$.
In this case, 
the width of the aliasing lobes is inversely proportional to the FOV
and the angular density compensation, $\dcft$, is uniform
(Eq.~\ref{eq:2d_dcf}).  See Section~\ref{sec:discussion} for more
discussion of the dual shapes.
Reducing the resolution also reduces the number of projections required.
For the same FOV shapes with isotropic resolution, the trajectories in
Fig.~\ref{fig:2Dpsfs_kmax} require between 13 and 36\% less
projections, demonstrating the
trade-off that can be made between resolution in a given dimension and acquisition time.

Animations of how the PSF evolves as the algorithm progresses are
available at \url{http://www-mrsrl.stanford.edu/~peder/radial_fovs}.
The principles of the sampling approximations used
(Fig.~\ref{fig:sampling_approx}) are visible in the movies.

\subsection{3D PSFs}
\label{sec:3DPSFs}
The PSFs for 3D cones trajectories are 2D PR PSFs rotated about one axis,
as is illustrated in Fig.~\ref{fig:3DPR_design}a and b.

\begin{figure}[t]
  \begin{center}
    \includegraphics[width=1.0\columnwidth]{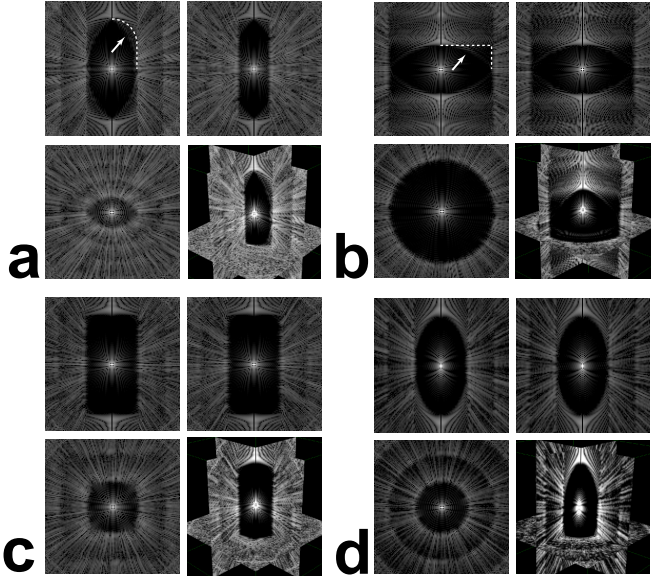}
    \caption{
      PSFs for 3D PR trajectories with anisotropic FOVs using the
      spiral-based design.
      FOVs: (a) Oval $\FOVt$ and ellipse $\FOVp$, (b) Cylinder, (c) Cuboid, and (d) Ellipsoid with
      an ellipsoid $\kmaxt$ shape, shrunk in the $k_z$ direction.  The top row for each
      shape shows the PSF in the $y=0$ and $x=0$ planes, and the bottom row shows the
      $z=0$ plane and a 3-plane view.
      Similarly to the 2D case, there is some low-level aliasing (arrows)
      inside the desired FOV shapes (dashed lines), which is
      particularly visible in (b).
    }
    \label{fig:3Dpsfs}
  \end{center}
\end{figure}

Figure~\ref{fig:3Dpsfs} shows some PSFs for 
spiral-based 3D PR sampling patterns.  For the variety of shapes
and dimensions, the desired
FOV is achieved, and using an oval for $\FOVt$ and ellipse for $\FOVp$ (Fig.~\ref{fig:3Dpsfs}a)
results in three different FOV dimensions.  
There are low-level aliasing artifacts present in these PSFs (arrows
in Fig.~\ref{fig:3Dpsfs}a and b). 
The cuboid
(Fig.~\ref{fig:3Dpsfs}c) has sharp corners in all three dimensions,
and is also well-defined at non-zero $z$-values, demonstrating that
the azimuthal sampling aliasing shape, $\FOVp$, is invariant in $z$,
as assumed in Eq.~\ref{eq:3DFOV}.
The PSF of an ellipsoid FOV with an ellipsoid $\kmaxt$ is
shown in Fig.~\ref{fig:3Dpsfs}d, demonstrating that variable $\kmaxt$ is
feasible.  This trajectory has a 50\% reduction in Z-gradient
strength requirements compared to the other trajectories because $\kmaxt$
limits the projection extents in $k_z$.  This halves the resolution
in $z$, but requires 33\% less projections than the same FOV shape with
1 pixel isotropic resolution.

\begin{figure}
  \begin{center}
    \includegraphics[width=\columnwidth]{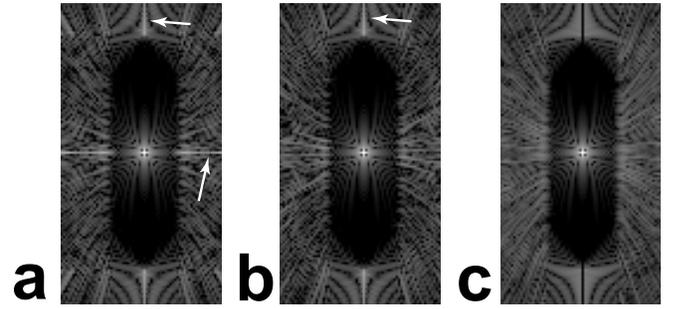}
    \caption{
      PSFs for various 3D PR design methods.      
      (a) Cones-based design without randomized $\phi_0$.
      (b) Cones-based design with randomized $\phi_0$.
      (c) Spiral-based design.
      The $x=0$ plane for the shape in Fig.~\ref{fig:3Dpsfs}a is
      shown, and the images are identically windowed.
      The arrows indicate coherent streaking artifacts, all of which
      are eliminated when using the spiral-based design.
    }
    \label{fig:coherent_aliasing}
  \end{center}
\end{figure}

Coherent aliasing lines are introduced when using the cones-based
design, as shown in Fig.~\ref{fig:coherent_aliasing}.
The use of a random $\phi_0$ in the cones-based design eliminates the
coherent line in the $x$-$y$ plane.  A line along
the $z$-axis remains because the polar projection spacings
around $\theta = \hpi$ are all identical.  
The tilt of the spiral in
$k_z$ diffuses the line in the spiral-based design.
These coherent lines would not affect fully-sampled acquisitions.
In undersampled applications, which take advantage of the diffuse 3D PR aliasing,
this could cause artifacts, making the spiral-based design advantageous.

\begin{figure}
  \begin{center}
    \includegraphics[width=0.9\columnwidth]{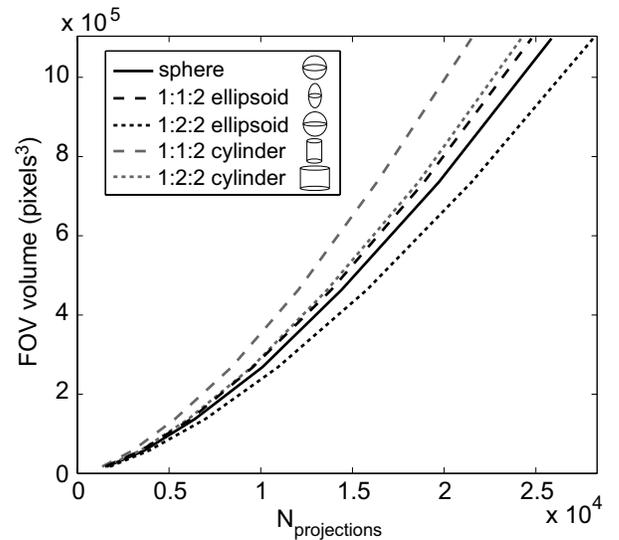}
    \caption{
      Number of projections for 3D anistropic FOV shapes, with ratios
      of $x$:$y$:$z$ lengths, versus the shape volume.  
      All FOVs have 1 pixel isotropic resolution, and 
      the curves shown are for both the cones-based and spiral-based
      design methods.
      Most of the anisotropic shapes are actually more efficient than
      the sphere, but at the cost of low-level aliasing within the FOV, seen in Fig.~\ref{fig:3Dpsfs}.
    }
    \label{fig:vol}
  \end{center}
\end{figure}

The scan volume efficiency for some 3D anisotropic FOV shapes is shown in
Fig.~\ref{fig:vol}.  The anisotropic shapes are generally more
efficient, although there is more variation in efficiency than in the 2D case
(Fig.~\ref{fig:area}).  The increased efficiency is at the expense of
some low-level aliasing inside the FOV.
The 1:1:2 ($x$:$y$:$z$) shapes only have anisotropy in the polar sampling, and they
are more efficient than the 1:2:2 shapes that also have azimuthal
sampling anisotropy.  This difference is caused by
the anisotropic $\FOVp$ cutting off portions of $\FOVt$, which can be
seen in Fig.~\ref{fig:3DPR_design}. 

\begin{figure}
  \begin{center}
    \includegraphics[width=.8\columnwidth]{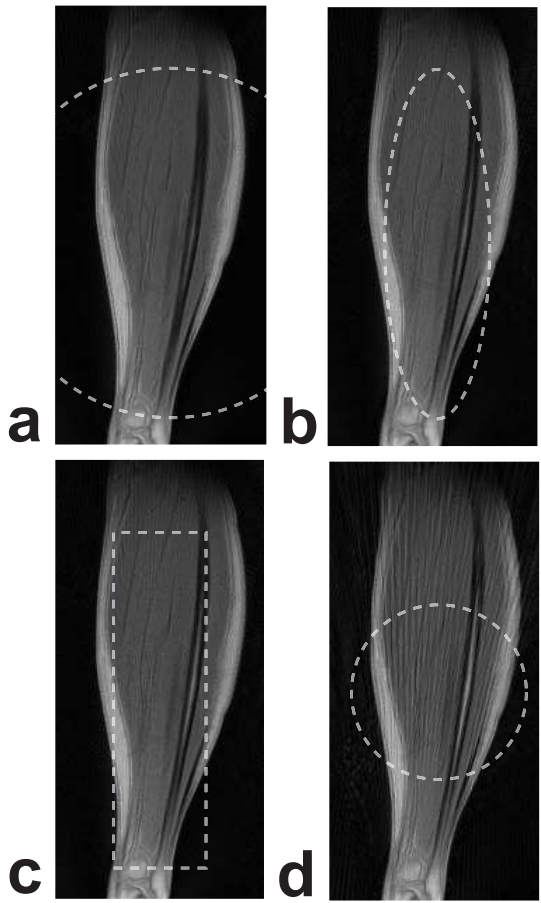}
    \caption{
      2D PR Lower leg images with various FOV shapes, shown by the
      dashed lines.
      (a) 25 cm circular FOV requiring 393 projections, 
      (b) 7.5 x 25 cm elliptical FOV requiring 197 projections, 
      (c) 6.5 x 24 cm rectangular FOV requiring 195 projections, 
      (d) 12.5 cm circular FOV requiring 196 projections.  
      The non-circular FOV images (b,c) show no increase in aliasing artifacts or
      loss of resolution but were acquired with half the projections.  The
      circular FOV acquired with half the projections (d) results in
      significant aliasing artifacts.
    }
    \label{fig:leg_images}
  \end{center}
\end{figure}

\subsection{MRI Experiments}
In vivo 2D PR leg images acquired with isotropic and anisotropic FOVs
are shown in Fig.~\ref{fig:leg_images}.  The reduced FOV images (b-d)
were acquired with half the number of projections as the full FOV
image (a).  The isotropic reduced FOV (d) results in significant
streaking aliasing artifacts, while using anisotropic reduced FOVs
tailored to the shape of the leg (b,c) results in no increase in
artifact compared to the full isotropic FOV image. 
The images without aliasing (a-c) are all slightly undersampled, as shown by the
overlaid FOV shapes, but no aliasing is visible due to the
relatively diffuse aliasing pattern of PR.

\begin{figure}[h]
  \begin{center}
    \includegraphics[width=.8\columnwidth]{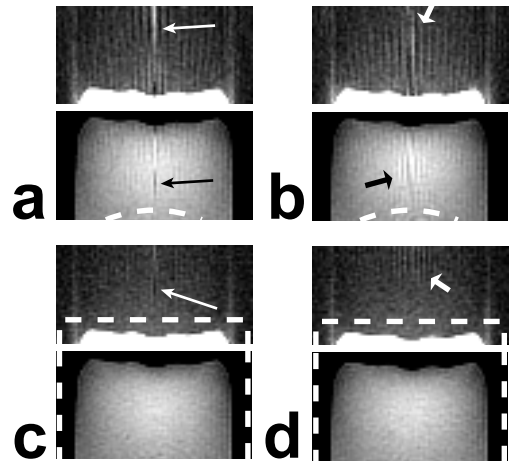}
    \caption{
      3D PR phantom images, all acquired with the same 
      number of projections.
      Isotropic FOV using the (a) cones-based method and 
      (b) spiral-based method.
      Cylindrical FOV using the (c) cones-based method and 
      (d) spiral-based method.
      The top images are all windowed identically to show the 
      aliasing artifacts outside the bottle (white arrows). 
      The bottom images are also identically windowed to show
      the aliasing artifacts within the bottle for the 
      undersampled, isotropic FOVs (black arrows).
      The dashed lines indicate the supported FOV.
    }
    \label{fig:bottle}
  \end{center}
\end{figure}

Figure~\ref{fig:bottle} shows a representative slice from 3D PR
acquisitions of a water bottle phantom using different FOVs,
each with the same number of projections.  In the isotropic FOV acquisitions,
streaking artifacts are visible
within the bottle and emanate from the edge of the bottle
(arrows in Fig.~\ref{fig:bottle}a and b)
because the FOV is not large enough.  
These artifacts have a peak amplitude of 23.2\% of the signal 
for the cones-based design and 11.5\% for the spiral-based design.
By using a
fully-sampled cylindrical FOV that matches the bottle's shape, 
these streaks are completely shifted outside of the bottle. 
(arrows in Fig.~\ref{fig:bottle}c and d).
The differences in aliasing diffusivity between the 
two 3D PR design methods, shown in the PSFs in
Fig.~\ref{fig:coherent_aliasing}, can also be seen in the images.
There is a single, prominent aliasing streak when using the cones-based design
(long, thin arrows), which is diffused with
the spiral-based design (short, fat arrows), and the peak 
aliasing amplitudes also reflect this difference.

\begin{figure*}
  \begin{center}
    \includegraphics[width=1.8\columnwidth]{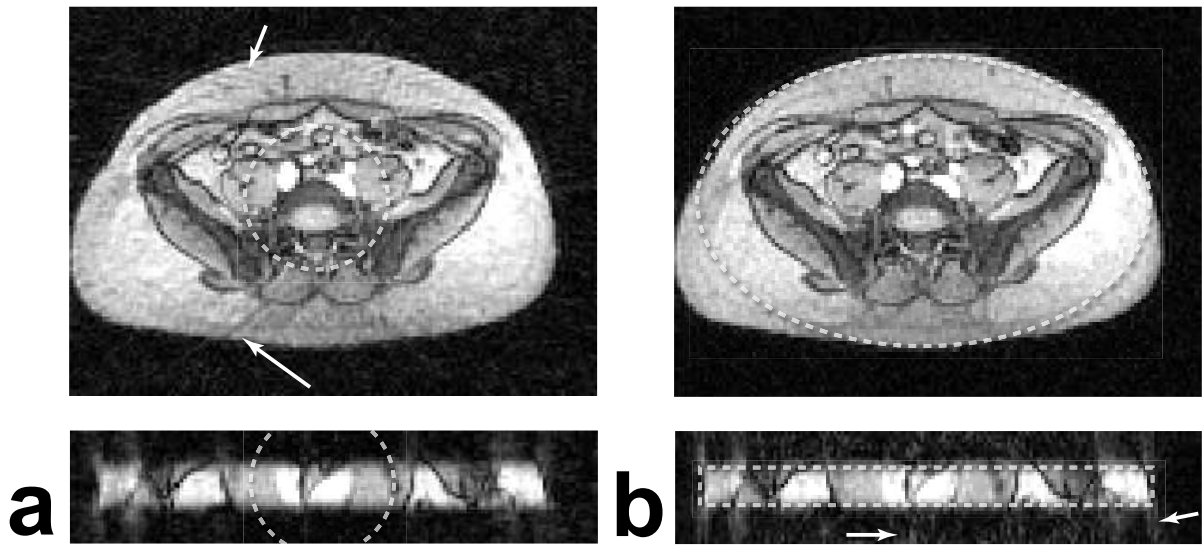}
    \caption{
      Axial (top) and coronal (bottom) slices from a thin-slab 3D PR
      abdomen data-set with 3 mm isotropic resolution acquired in a single breath-hold.
      (a) Isotropic 11.4 cm FOV requiring 2303 projections.
      (b) Cylindrical 36 x 23 x 3 cm FOV requiring 2368 projections.
      Both trajectories were designed using the spiral-based method.
      Streaking artifacts result from the high axial undersampling of the isotropic FOV
      (arrows in (a)).  The tailored FOV eliminates these artifacts
      by reducing FOV in the slab dimension.  This leads to the streaks in the
      coronal slice (arrows in (b)), but these are outside of the imaging volume.
    }
    \label{fig:abdomen}
  \end{center}
\end{figure*}

\begin{figure*}
  \begin{center}
    \includegraphics[width=1.8\columnwidth]{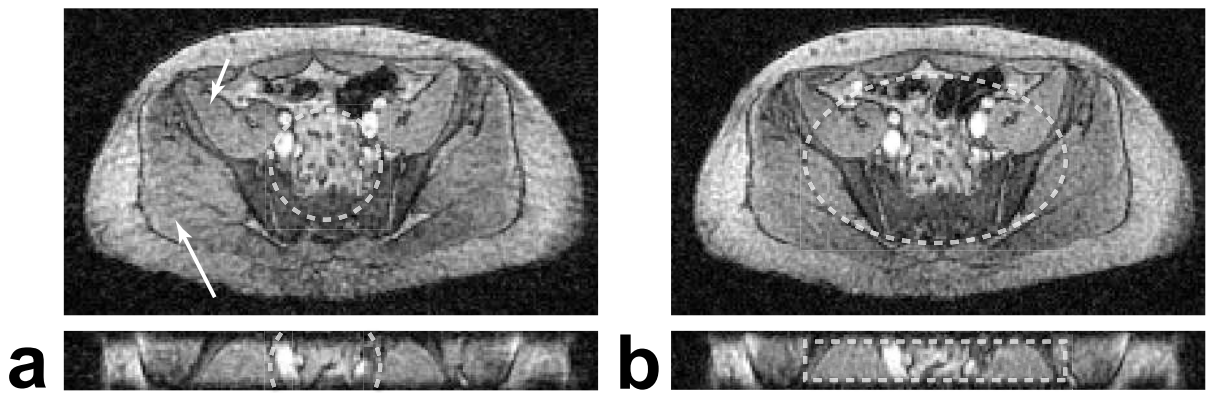}
    \caption{
      Same as Fig.~\ref{fig:abdomen}, except with 2 mm isotropic
      resolution and increased undersampling.
      (a) Isotropic 8 cm FOV requiring 2519 projections.
      (b) Cylindrical 19.6 x 12.2 x 2.8 cm FOV requiring 2529 projections.
      The undersampled cylindrical FOV has no noticeable streaking
      artifacts (arrows), which obscure the anatomy with the isotropic FOV.
    }
    \label{fig:abdomen_2mm}
  \end{center}
\end{figure*}

In vivo 3D PR images of a 3 cm abdomen slab acquired in a single breath-hold 
are shown in Figs.~\ref{fig:abdomen} and~\ref{fig:abdomen_2mm}.
The isotropic FOV images have significant streaking artifacts within
the abdomen (arrows in Fig.~\ref{fig:abdomen}a and~\ref{fig:abdomen_2mm}a)
because of the high undersampling ratios.
Using a thin, squished cylindrical FOV tailored to the anatomy and the
excited slab eliminates these streaking artifacts, as shown in 
Fig.~\ref{fig:abdomen}b.  A small degree of undersampling in this
tailored FOV is also
tolerated, as evidenced by the lack of streaking artifacts in
Fig.~\ref{fig:abdomen_2mm}b.
The isotropic FOV streaking also results in a higher level of signal
outside of the body in the axial images.
More streaking is seen just outside the slab in the coronal images with 
the anisotropic FOV (arrows in Fig.~\ref{fig:abdomen}b).
This is expected because of the smaller superior-inferior
dimension with the anisotropic FOV relative to the spherical FOV, for which the streaks in this
dimension are outside of the displayed region.
Acquiring a fully-sampled 36 cm isotropic FOV at 3 mm isotropic resolution
requires 22656 projections, while an undersampled 19.6 cm FOV at 2 mm
resolution requires 14744 projections, both of which are 
prohibitively large for single breath-hold imaging.

\section{Discussion}
\label{sec:discussion}
The algorithms presented are designed with fully-sampled FOV
parameters, but will also be beneficial for
undersampled radial imaging applications, as demonstrated in
Fig.~\ref{fig:abdomen_2mm}.  For undersampled applications like
VIPR~\cite{BargerVIPR}, the desired FOV shape and size should be a
scaled down version of the region-of-interest for equivalent
undersampling factors in every dimension.
The anisotropic FOV projections can also be applied to the highly constrained
backprojection (HYPR) method for time-resolved
MRI~\cite{MistrettaHYPR} because they can be reconstructed by filtered backprojection.  
Exam-specific FOVs can be used in all applications because
the computation time of the design algorithms is small.
Using an initial localizing scan, the
desired FOV could be drawn or automatically detected and the
acquisition tailored appropriately.

For some FOV and $\kmax$ shapes, the resulting PSFs have
low-level aliasing within the desired FOV.  This is due to the finite
width of the aliasing lobes, as shown in Fig.~\ref{fig:sampling_approx}b, which is not
accounted for in our design algorithms.  The projection angles are
designed based on the peak at the center of these aliasing lobes,
leaving the potential for up to half of the lobe to overlap inside the FOV.
Overlap is more likely to occur as the radial separation of neighboring
lobes increases, and thus is correlated with
$\left | \frac{d\FOV(\phi)}{d\phi} \right |$.
The online animations of the PSFs show how the
origin of the low-level aliasing is primarily along the longer edges
where this derivative is the largest.
This aliasing 
does not originate from minima or maxima of the FOV where the
angular derivative is zero. 

Narrower shapes, which have a larger angular derivative, have more desirable efficiency
curves in Fig.~\ref{fig:area} because they have more
low-level aliasing within the FOV.
For the shapes in Fig.~\ref{fig:area}, the total low-level aliasing power is approximately
0.60 and 1.3\% of the main lobe power for the 1.5:1 and 2:1 ellipses,
respectively, and 0.67, 0.91, and 1.2\% for the 1:1, 1.5:1, and 2:1 rectangles.
This confirms that narrower shapes have more low-level aliasing and
that, in general, this aliasing is small.

There is no low-level aliasing in the dual shape sampling patterns seen in
Fig.~\ref{fig:2Dpsfs_kmax}b, c, and d.  
This suggests that the low-level aliasing may be more correctly correlated with 
the derivative of $\FOV(\phi) / \kmax(\phi+\hpi)$,
which is zero for the dual shapes.  This derivative is very large for
the FOV and $\kmax$ shapes in 
Fig.~\ref{fig:2Dpsfs_kmax}e, which has significant aliasing inside
the desired FOV.
Another observation is that
dual ellipse sampling can be formed by scaling an isotropic 
sampling pattern along one axis.  By Fourier theory, this transformation
scales the PSF along the corresponding axis by the inverse
and introduces no low-level aliasing.
It is possible that the other shapes may be formed by some
transformation of isotropic case. 
We have found no derivation confirming this suggested low-level
aliasing correlation or any general transformation starting with isotropic sampling.

With our method, it is possible to use 3D radial imaging trajectories
to image thin slabs with isotropic resolution, as demonstrated in
Figs.~\ref{fig:abdomen} and~\ref{fig:abdomen_2mm}.  Previously, this
was achieved by using stacks of projections~\cite{PetersPRMRA,
  VigenPRTRICKS}, which lacks some of the benefits of a true 3D radial
acquisition, such as ultra-short TEs.
Stacking the projections also results in a coherent aliasing streak
along the stacking direction, similar to the streak resulting from the
cones-based design (see Fig.~\ref{fig:coherent_aliasing}), which may be 
related to artifacts reported
in the stacking dimension with contrast enhancement~\cite{BargerVIPR}.
Both methods support one dimension of variable resolution, and
stacks of 2D anisotropic FOV projections are also possible.

This projection design can also be applied to other centric-based
k-space trajectories.  It can be directly applied to twisting radial-line
(TwiRL) trajectories \cite{JacksonTwirl} by specifying the angular
spacing of the different acquisitions.  Anisotropic resolution can
also be incorporated by adjusting the twist in
the radial line to match $\kmax(\phi)$.
Interleaved spiral acquisitions can also be adjusted for anisotropic FOVs.  
The angular spacing of adjacent interleaves can be
determined by our algorithm, and spirals with many interleaves will
benefit the most from this adjustment.
This could also be combined with the
previous anisotropic FOV spiral method \cite{KingFOV} which describes
the design of the spirals themselves.





\section{Conclusion}
We have introduced a new method that designs projections for anisotropic FOVs in
radial imaging.  These FOVs can be precisely tailored to
non-circular objects or regions-of-interest in 2D and 3D imaging.
This allows for scan time reductions without introducing
aliasing artifacts.  For undersampled applications, this method
allows for reduction of aliasing artifacts.

Algorithms have been presented for the design of 2D and 3D PR
trajectories, as well as 3D cones.
There is no loss in FOV area efficiency when using anisotropic FOVs.
The algorithms are very simple and fast, allowing them to be computed
on-the-fly.  They also support variable trajectory extents which can
be used in MRI to favor certain gradients or relax gradient
constraints in 3D cones.



\appendices


\section{Full-projection Spiral-based Design}
\label{app:spiral_full}
For a spiral-based full-projection design, the spiralling path is 
sampled for $0 \le \theta \le \hpi$ by using $\thetawid = \hpi$ for the set of initial polar samples,
and interpolating to $\hT[N_{polar}+1] = \hpi$ and $\hKmax[N_{polar}+1] = \kmaxt(\hpi)$. 
Near $\theta = \hpi$, the
polar sample spacing is not as desired because the opposite
ends of the full-projections are spiralling in opposing directions.
The spacing between these opposing turns varies approximately linearly from
1.5 to 0.5 times the desired spacing over each of the final two
half-turns of the spiral.
This spacing discrepancy also occurs with the isotropic
3D PR spiral design for full-projections\cite{WongRoos}.
This undersampling can be compensated for by adding an 
extra quarter-turn to the spiral.  To do this, the interpolation is
carried out to additional samples of
$\hT[N_{polar}+2] = \hpi + \frac{1}{4 \kmaxp \FOVt(\pi)}$ and 
$\hKmax[N_{polar}+2] = \kmaxt(\hT[N_{polar}+2])$, 
with $N_{N_{polar}+1,\est} = \frac{N_{\phi,\est}}{4}$.

The density compensation must also be slightly modified to accomodate
both the opposing spiral paths and the extra quarter-turn.
Based on the 3D spiral geometry, it can be found that the total spacing
from adjacent turns decreases approximately linearly between 1 and 0.5 times the desired spacing
identically over the two last half-turns.
Thus, the dcf from Eq.~\ref{eq:3d_dcf} is weighted over each
of the half-turns separately by a linear ramp from 1 to 0.5 as $\theta$
increases.
\bibliographystyle{IEEEtran}
\bibliography{pederrefs,IEEEabrv}

\begin{thebibliography}{10}
\providecommand{\url}[1]{#1}
\csname url@rmstyle\endcsname
\providecommand{\newblock}{\relax}
\providecommand{\bibinfo}[2]{#2}
\providecommand\BIBentrySTDinterwordspacing{\spaceskip=0pt\relax}
\providecommand\BIBentryALTinterwordstretchfactor{4}
\providecommand\BIBentryALTinterwordspacing{\spaceskip=\fontdimen2\font plus
\BIBentryALTinterwordstretchfactor\fontdimen3\font minus
  \fontdimen4\font\relax}
\providecommand\BIBforeignlanguage[2]{{%
\expandafter\ifx\csname l@#1\endcsname\relax
\typeout{** WARNING: IEEEtran.bst: No hyphenation pattern has been}%
\typeout{** loaded for the language `#1'. Using the pattern for}%
\typeout{** the default language instead.}%
\else
\language=\csname l@#1\endcsname
\fi
#2}}

\bibitem{Lauterbur}
P.~C. Lauterbur, ``Image formation by induced local interactions: Examples
  employing nuclear magnetic resonance,'' \emph{Nature}, vol. 242, pp.
  190--191, May 1973.

\bibitem{IrarrazabalCones}
P.~Irarrazabal and D.~G. Nishimura, ``Fast three dimensional magnetic resonance
  imaging,'' \emph{Magn\ Reson\ Med}, vol.~33, no.~5, p. 656, May 1995.

\bibitem{GurneyCones}
P.~T. Gurney, B.~A. Hargreaves, and D.~G. Nishimura, ``Design and analysis of a
  practical {3D} cones trajectory,'' \emph{Magn\ Reson\ Med}, vol.~55, no.~3,
  pp. 575--582, Mar. 2006.

\bibitem{GloverPR}
G.~H. Glover and J.~M. Pauly, ``Motion artifact reduction with projection
  reconstruction imaging,'' \emph{Magn\ Reson\ Med}, vol.~28, no.~2, pp.
  275--289, Dec. 1992.

\bibitem{NishimuraFlow}
D.~Nishimura, J.~I. Jackson, and J.~M. Pauly, ``On the nature and reduction of
  the displacement artifact in flow images,'' \emph{Magn\ Reson\ Med}, vol.~22,
  no.~2, pp. 481--492, Dec. 1991.

\bibitem{PetersPRMRA}
D.~C. Peters, F.~R. Korosec, T.~M. Grist, W.~F. Block, J.~E. Holden, K.~K.
  Vigen, and C.~A. Mistretta, ``Undersampled projection reconstruction applied
  to {MR} angiography,'' \emph{Magn\ Reson\ Med}, vol.~43, no.~1, pp. 91--101,
  Jan. 2000.

\bibitem{BargerVIPR}
A.~V. Barger, W.~F. Block, Y.~Toropov, T.~M. Grist, and C.~A. Mistretta,
  ``Time-resolved contrast-enhanced imaging with isotropic resolution and broad
  coverage using an undersampled {3D} projection trajectory.'' \emph{Magn\
  Reson\ Med}, vol.~48, no.~2, pp. 297--305, Aug. 2002.

\bibitem{LuMultiechoPR}
A.~Lu, E.~Brodsky, T.~M. Grist, and W.~F. Block, ``Rapid fat-suppressed
  isotropic steady-state free precession imaging using true {3D}
  multiple-half-echo projection reconstruction,'' \emph{Magn\ Reson\ Med},
  vol.~53, no.~3, pp. 692--699, Mar. 2005.

\bibitem{VigenPRTRICKS}
K.~K. Vigen, D.~C. Peters, T.~M. Grist, W.~F. Block, and C.~A. Mistretta,
  ``Undersampled projection-reconstruction imaging for time-resolved
  contrast-enhanced imaging,'' \emph{Magn\ Reson\ Med}, vol.~43, no.~2, pp.
  170--176, Feb. 2000.

\bibitem{StehningCoronaryFFE}
C.~Stehning, P.~B\"{o}rnert, K.~Nehrke, H.~Eggers, and O.~D\"{o}ssel, ``Fast
  isotropic volumetric coronary {MR} angiography using free-breathing {3D}
  radial balanced {FFE} acquisition,'' \emph{Magn\ Reson\ Med}, vol.~52, no.~1,
  pp. 197--203, July 2004.

\bibitem{Stehning3DPRNav}
------, ``Free-breathing whole-heart coronary {MRA} with {3D} radial {SSFP} and
  self-navigated image reconstruction,'' \emph{Magn\ Reson\ Med}, vol.~54,
  no.~2, pp. 476--480, Aug. 2005.

\bibitem{GateByd}
P.~D. Gatehouse and G.~M. Bydder, ``Magnetic resonance imaging of short {T2}
  components in tissue,'' \emph{Clin\ Radiol}, vol.~58, no.~1, pp. 1--19, Jan.
  2003.

\bibitem{RobsonJCAT}
M.~D. Robson, P.~D. Gatehouse, M.~Bydder, and G.~M. Bydder,
  ``\BIBforeignlanguage{eng}{Magnetic resonance: an introduction to ultrashort
  {TE (UTE)} imaging},'' \emph{\BIBforeignlanguage{eng}{J\ Comput\ Assist\
  Tomogr}}, vol.~27, no.~6, pp. 825--846, Nov./Dec. 2003.

\bibitem{Rahmer3DUTE}
J.~Rahmer, P.~B\"{o}rnert, J.~Groen, and C.~Bos, ``Three-dimensional radial
  ultrashort echo-time imaging with {T2} adapted sampling,'' \emph{Magn\ Reson\
  Med}, vol.~55, no.~5, pp. 1075--1082, May 2006.

\bibitem{LarsonT2supp}
P.~E.~Z. Larson, P.~T. Gurney, K.~Nayak, G.~E. Gold, J.~M. Pauly, and D.~G.
  Nishimura, ``Designing long-{T2} suppression pulses for ultrashort echo time
  imaging,'' \emph{Magn\ Reson\ Med}, vol.~56, no.~1, pp. 94--103, July 2006.

\bibitem{GuPC-VIPR}
T.~Gu, F.~R. Korosec, W.~F. Block, S.~B. Fain, Q.~Turk, D.~Lum, Y.~Zhou, T.~M.
  Grist, V.~Haughton, and C.~A. Mistretta, ``{PC VIPR}: A high-speed {3D}
  phase-contrast method for flow quantification and high-resolution
  angiography,'' \emph{Am\ J\ Neuroradiol}, vol.~26, no.~4, pp. 743--749, Apr.
  2005.

\bibitem{SchefflerNUPR}
K.~Scheffler, ``Tomographic imaging with nonuniform angular sampling,''
  \emph{J\ Comput\ Assist\ Tomogr}, vol.~23, no.~1, pp. 162--166, Jan./Feb.
  1999.

\bibitem{SchefflerRFOV}
K.~Scheffler and J.~Hennig, ``Reduced circular field-of-view imaging,''
  \emph{Magn\ Reson\ Med}, vol.~40, no.~3, pp. 474--480, Sept. 1998.

\bibitem{LauzonPolarSampling}
M.~L. Lauzon and B.~K. Rutt, ``Effects of polar sampling in k-space,''
  \emph{Magn\ Reson\ Med}, vol.~36, no.~6, pp. 940--949, Dec. 1996.

\bibitem{GloverBoron}
G.~H. Glover, J.~M. Pauly, and K.~M. Bradshaw, ``Imaging {$^{11}$B} with a {3D}
  projection reconstruction method,'' \emph{J\ Magn\ Reson\ Imaging}, vol.~2,
  no.~1, pp. 47--52, Jan./Feb. 1992.

\bibitem{WongRoos}
S.~T.~S. Wong and M.~S. Roos, ``A strategy for sampling on a sphere applied to
  {3D} selective {RF} pulse design,'' \emph{Magn\ Reson\ Med}, vol.~32, no.~6,
  pp. 778--784, Dec. 1994.

\bibitem{BeattyGridding}
P.~J. Beatty, D.~G. Nishimura, and J.~M. Pauly, ``Rapid gridding reconstruction
  with a minimal oversampling ratio,'' \emph{IEEE Trans\ Med\ Imaging},
  vol.~24, no.~6, pp. 799--808, June 2005.

\bibitem{TsaiVdens}
C.-M. Tsai and D.~Nishimura, ``Reduced aliasing artifacts using
  variable-density $k$-space sampling trajectories,'' \emph{Magn\ Reson\ Med},
  vol.~43, no.~3, pp. 452--458, Mar. 2000.

\bibitem{MistrettaHYPR}
C.~A. Mistretta, O.~Wieben, J.~Velikina, W.~Block, J.~Perry, Y.~Wu, K.~Johnson,
  and Y.~Wu, ``Highly constrained backprojection for time-resolved {MRI},''
  \emph{Magn\ Reson\ Med}, vol.~55, no.~1, pp. 30--40, Jan. 2006.

\bibitem{JacksonTwirl}
J.~I. Jackson, D.~G. Nishimura, and A.~Macovski, ``Twisting radial lines with
  application to robust magnetic resonance imaging of irregular flow,''
  \emph{Magn\ Reson\ Med}, vol.~25, no.~1, pp. 128--139, May 1992.

\bibitem{KingFOV}
K.~F. King, ``\BIBforeignlanguage{eng}{Spiral scanning with anisotropic field
  of view},'' \emph{\BIBforeignlanguage{eng}{Magn\ Reson\ Med}}, vol.~39,
  no.~3, pp. 448--456, Mar. 1998.

\end{thebibliography}



\end{document}